\documentclass[twocolumn,showpacs,amsmath,amssymb,prd,nofootinbib]{revtex4}
\usepackage{epsfig}
\def\sss{\scriptscriptstyle}
\def\^#1{^{\sss #1}}
\def\_#1{_{\sss #1}}
\def\beq{\begin{equation}}
\def\eeqno#1{\label{#1}\end{equation}}

\def\kms{~{\rm km~s^{-1}}}

\def\kpc{~{\rm Kpc}}
\def\mpc{~{\rm Mpc}}

\def\msun{M\_{\odot}}
\def\az{a\_{0}}

\def\l0{\ell\_{0}}

\def\l{\lambda}

\def\k{\kappa}

\def\e{\eta}

\def\b{\beta}

\def\vv{{\bf v}}

\def\gN{g\_N}

\def\av#1{\langle #1\rangle}

\def\tot{{2\over 3}}

\def\suK{~\msun/L\_{K,\odot}}
\def\suV{~\msun/L\_{V,\odot}}
\def\sigv{\sigma}
\def\sigm{\sigma\_{\small M}}
\def\Rmk{Ref. \cite{mk11}}

\begin{document}
\title{MOND in galaxy groups: A superior sample}
\author{Mordehai Milgrom }
\affiliation{Department of Particle Physics and Astrophysics, Weizmann Institute, Rehovot Israel 7610001}

\begin{abstract}
Intermediate-richness galaxy groups are an important testing ground for MOND. First, they constitute a distinct type of galactic systems, with their own evolution histories and underlying physical processes; second, they probe little-chartered regions of parameter space, as they have baryonic masses similar to massive galaxies, and similar velocity dispersions, but much larger sizes -- similar to cluster cores (or even to clusters), but much lower dispersions. Importantly in the context of MOND, they have the lowest internal accelerations reachable inside galactic systems. Following my recent analysis of MOND in galaxy groups, I came across a much superior sample, which I analyze here. This extensive catalog permits strict quality cuts that still leave a large sample of 56 medium-richness groups, better suited for dynamical analysis -- e.g., in having a large number ($\ge 15$) of members with measured velocities.
I find that these groups obey the deep-MOND relation between baryonic mass, $M\_M$, and velocity dispersion, $\sigv$:
$M\_MG\az=(81/4)\sigv^4$, with individual, MOND, mass-to-light ratios, $M\_M/L\_K$ of order $1\suK$, and a sample median value of $(M\_M/L\_K)_{med}=0.7\suK$. These compare well with stellar values deduced for single galaxies, and with values deduced from population-synthesis analyses. In contrast, the dynamical, Newtonian $M_d/L\_K$ values are much larger -- several tens solar units, and $(M_d/L\_K)_{med}=37\suK$. The same MOND relation describes (isolated) dwarf spheroidals -- two-three orders smaller in size, and seven-eight orders lower in mass. The groups conformation to the MOND relation is equivalent to their lying on the deep-MOND branch of the ``mass-discrepancy-acceleration relation'', $g\approx (\gN\az)^{1/2}$, for $g$ as low as a few percents of $\az$ ($\gN$ is the Newtonian, baryonic, gravitational acceleration, and $g$ the actual one). This argues against systematic departure from MOND at extremely low accelerations (as has been suggested).
This conformation also argues against the hypothesis that the remaining MOND conundrum in cluster cores bespeaks a breakdown of MOND on large-distance scales; our groups are as large as cluster cores, but do not show obvious disagreement with MOND.
I also discuss the possible presence of the idiosyncratic, MOND external-field effect.

\end{abstract}
\maketitle

\section{Introduction}
Galaxy groups are important in testing MOND (\cite{milgrom83}, reviewed, e.g., in Refs. \cite{fm12,milgrom14c}) -- especially {\it vis \`{a} vis} the dark-matter paradigm. Such groups are a class of galactic objects distinct from others, such as dwarf spheroidals, massive galaxies, and galaxy clusters, with different histories of formation and evolution, different physical processes that affect their evolution, and possibly a different dynamical state. Such differences are expected to lead to disparate
relations between baryons and dark-matter in the dark-matter paradigm, with strongly history-dependent present states (see e.g., Refs. \cite{kroupa16,ok18}). In contradistinction,  MOND predicts strict relations between the baryons and the observed dynamics that are oblivious to history, as long as the system under study, be it a dwarf satellite, a galaxy, or a galaxy group, is in virial equilibrium.
\par
Intermediate-richness groups (as distinct from galaxy clusters) also probe a different region of parameter space from what can be reached with other galactic systems. They have masses and internal velocities similar to massive galaxies, but are tens of times larger.
They are similar in size to galaxy-cluster cores, but have much lower velocity dispersions. They thus have much lower internal accelerations than found in either galaxies or galaxy clusters. In fact -- importantly in the context of MOND -- their internal acceleration are the lowest accessible in galactic systems, typically several times lower even than is typical in dwarf spheroidal satellites.\footnote{Such low accelerations have been probed in MOND, far outside galaxies, using weak gravitational lensing \cite{milgrom13,browuer17}, but only statistically for samples of galaxies, not individually.} I will find no departure from the predictions of MOND in such groups, which argues against breakdown of MOND at extremely low accelerations -- a possibility that was raised e.g., in Ref. \cite{lelli17} based on analysis of ultra-faint dwarf galaxies.
\par
Because such groups are as large as cluster cores, or even as clusters, they help us pinpoint the reasons for the remaining MOND conundrum in clusters, and whether it is due to some breakdown of MOND at large distances -- as has been suggested occasionally -- or to some other attribute of clusters. The fact that MOND works well for the present sample, as I will find, means that it is not a question of system size.
\par
In a recent paper (\cite{milgrom18}, hereafter Paper I) I analyzed MOND in small-to-intermediate-richness galaxy groups listed in the three compilations in Refs. \cite{KKK17,KNK15,KKN14}, totaling 53 groups. While an advance over previous MOND analyses of galaxy groups, this study still leaves much to be desired. For example, most of the groups analyzed have only a small number of members with measured velocities, with only eleven groups having more than ten measured velocities.
It is greatly advantageous to analyze groups with a large number of observed member velocities, in order to mitigate some of the various uncertainties that beset the identification and analysis of groups, which assumes that they are bound, virialized, isotropic, isolated systems. Indeed, Paper I found clear-cut agreement with the predictions of MOND for the groups with more than 10 measured velocities, while those with a small number of velocities show large scatter.
\par
In the meanwhile, I came across an earlier, much more extensive compilation \Rmk~ that identified 365 galaxy associations as ``galaxy groups''.
This large pool allows the application of  ``quality cuts'' that still leave a large subsample of groups better suited for dynamical analysis.
\par
Here, I use the subsample of groups in \Rmk~ with at least 15 members with measured velocities, and, furthermore, I leave out the (ten) groups at the very-rich end of the sample, to avoid having to account for the large contribution to the baryonic mass of hot gas these rich-groups/clusters are known to harbor.
I then repeat the analysis as described in Paper I for the remaining subsample of 56 groups.
\par
Galaxy groups were first analyzed in MOND in the very first MOND trilogy  \cite{milgrom83a}, where, however, only a small number of groups with reasonable number of observed radial velocities were available (only 5 would have pass our cut here).
Also, the deep-MOND virial relation we now use for analyzing groups was not known at the time.
In Ref. \cite{milgrom98}, I analyzed four group catalogs, but given only the published averages and median values of the luminosities and velocity dispersions for these catalogs, with no analysis of individual groups.
In Ref. \cite{milgrom02a}, I considered a small sample of 8 relatively nearby groups from Ref. \cite{tully}. Most of these groups have dynamical times comparable with the Hubble time; so it is questionable whether they are in virial equilibrium, and four of them had no more than 4 members with measured velocities.
Testing MOND in pressure-supported galactic systems, Ref. \cite{scarpa03} plots for the first time (in its Figure 7) the MOND ``acceleration-discrepancy relation'' -- also known as the ``mass-discrepancy-acceleration relation (MDAR) -- for such systems; it shows also some groups (with very large scatter).
All these analyses gave results consistent with MOND.
\par
In Sec. \ref{analysis}, I describe details of the analysis: the method used and the choice of subsample for analysis.
In Sec. \ref{results}, I describe the results, including a comparison with previous similar analysis of dwarf-spheroidal satellites of Andromeda. Section \ref{discussion} discusses the results and lists some known sources of uncertainties and scatter about the MOND relation.
\section{Analysis \label{analysis}}
\subsection{Method}
The MOND relation used to analyze the groups was described in detail, including the associated caveats, in Paper I.
Here I recap the method, and comment on additional aspects not discussed in Paper I. The caveats and known sources of systematics -- some of which are specific to the present sample -- are discussed in Sec. \ref{systematics}.
\par
The starting MOND relation is (\cite{milgrom14a} and references therein)
\beq\av{\av{(\vv-\vv_0)^2}}_t=\tot(MG\az)^{1/2}[1-\sum_i
(m_i/M)^{3/2}], \eeqno{nus}
 where $\vv$ is the three-dimensional velocity,
$\vv_0$ is the center-of-mass velocity,
 $\av{}$  is the mass-weighted average over the constituents, whose
masses are $m_i$, $\av{}_t$ is the long-time average, and $M$ is
the total mass.
Relation (\ref{nus}) applies to isolated systems (ideally of point masses), deep in the MOND regime.
\par
The groups I shall consider here are all very deep in the MOND regime. Assuming long-term stationarity -- which requires boundedness (but is not ensured by it) -- we replace the long-time average with the measured present-day value. The three-dimensional velocity dispersions is replaced by $\sqrt{3}\sigv$ , $\sigv$ being the line-of-sight component -- the only one that is measured. This assumes global velocity isotropy -- a strong assumption that is not valid, for example, if rotation is important.
For determining masses we use literature values of $\sigv$, which are not mass weighted ones, as appear in relation (\ref{nus}), and in any event are measured only from radial velocities of (usually small) subsample of system members.
\par
Since individual masses of all members are not known, one usually uses relation (\ref{nus}) for the case of a system made of $N\gg 1$ masses individually $\ll M$, in which case we get
\beq M\approx {81\over 4}\sigv^4(G\az)^{-1}.  \eeqno{nusa}
This is the approximate relation used in Paper I (and, e.g., in analysing the MOND dynamics of Andromeda satellites in Refs. \cite{mm13a,mm13b}; see also the application in Ref. \cite{durazo18} to the MOND analysis of elliptical galaxies). Here, I shall purposely consider only groups with at least 15 members with measured velocities (and possibly many more members); so the finite-$N$ correction is not large if, indeed, all masses are small compared with $M$.
But, $N\gg 1$ does not ensure the validity of relation (\ref{nusa}). For example, in the case of one very dominant mass, $M$, with all the rest consisting of ``test particles'', relation (\ref{nus}) reads, given the above approximation for the left-hand side,
\beq M\approx 9\sigv^4(G\az)^{-1},  \eeqno{nusus}
where $\sigv$ is the mass weighted dispersion of the test particles alone. This gives masses that are a factor of $9/4$ smaller than from eq. (\ref{nusa}) (or $\sigv$ values $\approx 20\%$ larger, for a given mass).
\par
In the present analysis, I use the sample of groups described in Sec. \ref{sample}, with group parameters as given in \Rmk, which I list in Table \ref{table1}. They are: the numbers of member galaxies with observed radial velocities,
 $N\_V$; the line-of-sight velocity dispersions, $\sigv$, deduced from the measured radial velocities; the harmonic radii, $R_h$; the K-band luminosities, $L\_K$; the Newtonian, dynamical masses, $M_d$; and the corresponding Newtonian mass-to-light ratios, $M_d/L\_K$.
For each group, I calculate from $\sigv$ the MOND baryonic mass, $M\_M$, using eq. (\ref{nusa}), and the corresponding mass-to-light ratio, $M\_M/L\_K$, to be compared with reasonable baryonic values. I also reverse the procedure and calculate the expected MOND velocity dispersion, $\sigm$, using eq. (\ref{nusa}), from the given luminosity and assuming a fiducial value of $M(baryon)/L\_K=1\suK$. All these are also given in Table \ref{table1}.

\subsection{Alternative forms \label{aform}}
The deep-MOND  $M-\sigv$ virial relations of the type discussed above are the proper MOND predictions to be tested in ``pressure-supported systems'' such as the groups. They do not require knowledge of sizes -- a direct result of the scale invariance of the deep-MOND limit \cite{milgrom09}.
Such relations can, however, be cast in a form that resembles the acceleration-discrepancy relation -- the basic MOND prediction for circular orbits in an axisymmetric field \cite{milgrom83,milgrom94}, as relevant to rotation curves of disc galaxies\footnote{Also known as the mass-discrepancy-acceleration relation (MDAR), aka ``radial-acceleration relation'' (RAR).} -- by introducing some characteristic system radius, $\bar R$, which is needed for the definition of accelerations.
\par
As explained in Ref. \cite{milgrom14}, despite their similar appearance, the global $M-\sigv$ relations are different and independent MOND predictions from the local, mass-asymptotic-speed
relation $V^4\_{\infty}=MG\az$. (For example, the former is valid only for systems wholly in the deep-MOND regime, the latter is valid for all systems.) The latter is, in fact, a corollary of the former for the case of a test mass on a circular orbit around a central (baryonic) mass $M$. The $M-V\_{\infty}$ relation is tantamount to the ``acceleration-discrepancy relation'' between the observed (MOND) acceleration $g=V^2\_{\infty}/R$, and the Newtonian acceleration $\gN=MG/R^2$: $g=(\az\gN)^{1/2}$, at all radii on the asymptotic rotation curve.
MOND has extended this relation also to the interiors of disc galaxies  at locations where $g\ll\az$.
\par
To write the $M-\sigv$ relation in terms of ``global'' acceleration parameters, take, as an estimate of the global Newtonian acceleration,
\beq \gN=M\_MG/(\k \bar R)^2 \eeqno{gngn}
 ($M\_M$ is the MOND mass as defined above, because it stands for the baryonic mass), and the global true (MOND) acceleration as \beq g=\b\sigv^2/\bar R, \eeqno{gsr}
 then, eq. (\ref{nusa}), for example, can be written as
\beq  g=\e(\az\gN)^{1/2},~~~~~~\e\equiv 2\b\k/9.  \eeqno{mdar}
 Since $\k$ and $\b$ have values of a few, $\e\approx 1$, and thus $g\approx(\az\gN)^{1/2}$.
 \par
Alternatively, we can write the ratio of the Newtonian to the baryonic (i.e., MOND) masses -- which is also a measure of the mass-, or acceleration-discrepancy -- in terms of $g$, as follows:
Take an estimate of the Newtonian, dynamical mass to be such that
\beq \frac{M_dG}{(\l\bar R)^2}=g=\frac{\b\sigv^2}{\bar R}, \eeqno{mdmd}
for some $\l$ of order 1. Then, with the above definitions
\beq  \frac{M_d}{M\_M}=\frac{4\b^2\l^2}{81}\frac{\az}{g}.  \eeqno{mmg}
Again, since $\b$ and $\l$ have values of a few, this translates to $M_d/M\_M\approx \az/g$.
I shall also compare the groups data with this relation.

\subsection{The sample \label{sample}}
Putative groups with a small number of observed velocities, $N\_V$, are not suitable for dynamical analysis; e.g., because they are likely to be chance, projected groupings and not bound, virialized groups (a caveat also pointed to in \Rmk); and because, even if the group itself is real, the resulting line-of-sight velocity dispersion may not be a good representation of the true, three-dimensional dispersion.
A large number of group members (of which $N\_V$ is some measure) is also needed to justify the use of eq. (\ref{nusa}).
\par
The catalog of 365 galaxy ensembles cataloged in \Rmk~as groups is large enough to allow quality cuts that still leave a large sample to be analyzed. Somewhat arbitrarily, I keep for analysis here only those groups with $N\_V\ge 15$, totaling 67 groups. (Seven groups of those considered in Paper I have $N\_V\ge 15$.)
These groups are all listed in Table \ref{table1} with the parameter values assigned to them in \Rmk~and the ones I calculated here.
\par
I also want to exclude from the analysis galaxy clusters and rich groups that contain large quantities of x-ray emitting hot gas.
The reason for this is twofold. First, if stars are not the dominant contribution to the baryonic mass, the $M/L$ ratios are not good indicators of the acceleration discrepancies, and the analysis must take into account the gas; but this is beyond the scope of the present analysis. Second, we already know from past analyses, for example in Ref. \cite{angus08}, that in such hot-gas-rich groups, MOND does not fully account for the discrepancies, and may require an additional baryonic contribution that we have not yet detected. This added so called ``cluster baryonic dark matter'', may in fact be part and parcel of the hot gas \cite{milgrom08}, since, to my knowledge, such remaining MOND discrepancies are only seen in systems with substantial amounts of hot gas.
\par
To bypass the need to check the individual groups systematically for the quantities of gas present, I -- again somewhat arbitrarily -- excluded from the final analysis the ten groups with $N\_V>50$.
\par
For example, the group around NGC 4472 (M49), with the largest $N\_V$ is deep in the virgo cluster.
The second group , ``NGC 3311'', is the Hydra I cluster (Abel 1060), whose baryons are dominated by hot gas \cite{richtler11,hilker18}. The group ``NGC 4696'' is Centaurus. The group ``NGC 3223'' is the Antlia cluster, which is known to be hot-gas dominated \cite{wong16}.
The group ``NGC 5044'' is also known to be gas rich, with gas dominating the baryonic mass; it is one of the groups studied in MOND in Ref. \cite{angus08}.
The group ``NGC 4261'' is also hot-gas dominated \cite{roussel00}.
\par
As seen in Table \ref{table1}, some of these rich groups show reasonable MOND $M\_M/L\_K$ values, but generally they have values considerably larger than $1 \suK$, which would be at least partly explained by the contributions of gas to the MOND baryonic mass.
\par
In addition, I excluded one more group, ``NGC 4216'', which \Rmk~ explicitly disqualifies (with several others that are not in my sample anyway) as having a very wrong assigned distance. I discuss and exemplify this issue in more detail in Sec. \ref{systematics}.
\par
All this leaves us with a sample of  56 groups (five groups of those considered in Paper I would remain after the second cut).
\par
While the ten rich groups/clusters, and ``NGC 4216'', with their resulting MOND quantities are shown in Table \ref{table1}, they are excluded from the plots in Sec. \ref{results}.
\section{Results \label{results}}
\begin{figure}[h]
	\centering
\includegraphics[width = 9.5cm] {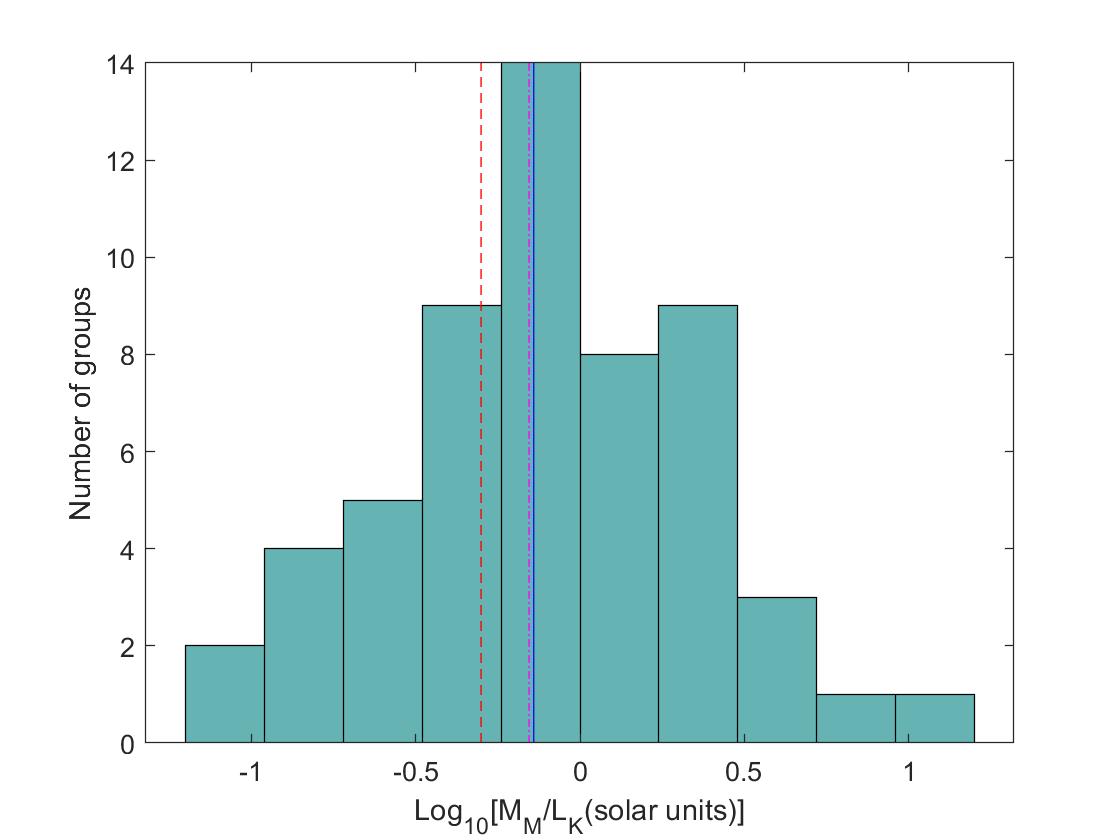}
\caption{The distribution of MOND dynamical $M\_M/L\_K\equiv (81/4G\az)(\sigv^4/L\_K)$ for our sample. The solid (blue) vertical line marks the sample median value of $0.72\suK$. The dash-dotted (magenta) line, near it, is the  value given in Ref. \cite{li18} for bulges (0.7), and the dashed (red) line is their value for discs (0.5), both deduced from MOND fits to rotation curves for the 3.6 micron photometric band. (See text for the small correction in comparing the two photometric bands involved.)}		\label{MONDMoL}
\end{figure}

\begin{figure}[h]
	\centering
\includegraphics[width = 9.5cm] {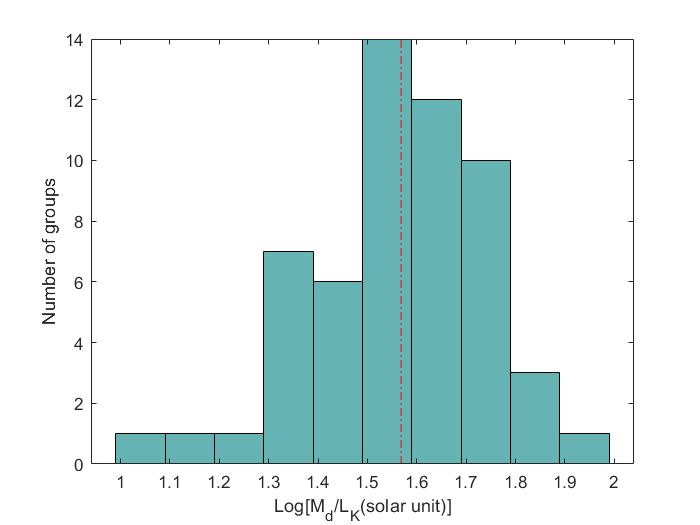}
\caption{The distribution of Newtonian, dynamical $M_d/L\_K$ ratios as derived in \Rmk~for the groups in our sample. The dash-dotted (red) vertical line marks the median value of $37\suK$.}		\label{NewtonianMoL}
\end{figure}

\begin{figure}[h]
	\centering
\includegraphics[width = 9.5cm] {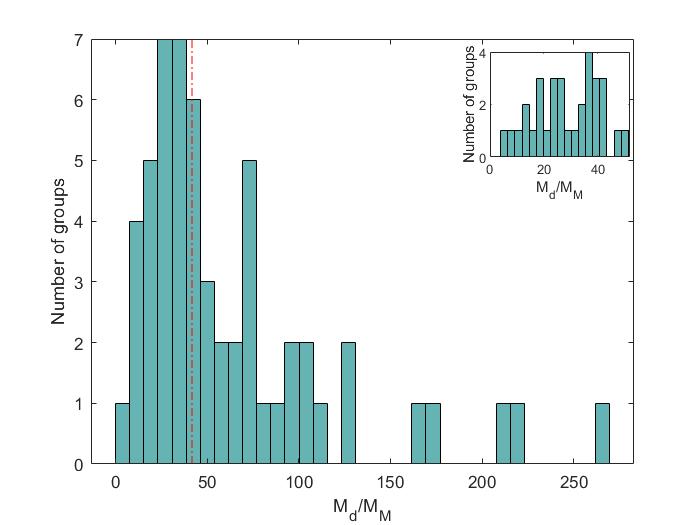}
\caption{The distribution of the Newtonian-to-MOND mass ratios $M_d/M\_M$, with the median value of 41 marked by the vertical dash-dotted line. The inset is a closeup for small values.}		\label{massratio}
\end{figure}

Using the $\sigv$ values given in \Rmk, I calculate the MOND masses of the groups, $M\_M$, from eq. (\ref{nusa}), and the resulting K-band mass-to-light ratio using the group K-luminosities from \Rmk.
The distribution of these $M\_M/L\_K$ values is shown in Fig. \ref{MONDMoL}, to be contrasted with the distribution of Newtonian, dynamical values $M_d/L\_K$ (from \Rmk), shown in Fig. \ref{NewtonianMoL}. While the Newtonian values are typically several tens solar units, and require the groups to be heavily dark-matter dominated in Newtonian dynamics, the MOND values are typically a factor of $40$ smaller, and fall around 1 solar unit, with a median value of $0.72\suK$ (compared with $37\suK$ for the Newtonian values). The distribution of $M_d/M\_M$ is shown in Fig. \ref{massratio}, with its median value of $(M_d/M\_M)\_{med}=41$.

\par
Inasmuch as the baryonic masses of the groups is dominated by stars, these MOND mass-to-light ratios should agree with what is known and what is expected of stellar $M/L\_K$ ratios, for example with what is deduced from population-synthesis models, or better yet, from the stellar $M/L\_K$ values directly deduced from rotation-curve analysis of individual galaxies.
\par
Indeed, the distribution of the baryonic, MOND $M\_M/L\_K$ values deduced here for the groups, and shown in Fig. \ref{MONDMoL}, is very similar  to those found for stellar values of discs and bulges in Ref. \cite{li18} (their Fig. 3) from rotation-curve analysis of many disc galaxies.\footnote{Reference \cite{li18} use the 3.6 micron band, not the K-band as here. Reference \cite{oh08} [their eq. (6) and Table 2] suggest correcting the K-band value down by about 10 percent to get the stellar value in the 3.6 micron band. This correction is anyhow small compared with the spread we find.}  The group distribution is centered at nearly the same values, but is somewhat wider, as expected from the sources of scatter that are hardly present in rotation-curve analysis (such as interlopers, anisotropies, etc. -- see Sec. \ref{systematics}). These MOND mass-to-light ratios also agree with earlier analyses of rotation curves (e.g., \cite{sv98}, their Fig. 2), and with model calculations based on population synthesis (e.g., Ref. \cite{bdj01}).
\par
It is worth reversing the procedure; namely, assume $M/L\_K=(M/L\_K)\_0$ for some reasonable reference value. Then calculate from eq. (\ref{nusa}) the MOND-predicted values, $\sigm$, from $L\_K$,
to be compared with the measured $\sigv$. I do this with $(M/L\_K)\_0=1\suK$,  and show $\sigm$ in Table \ref{table1}, together with $\sigm/\sigv$.
By the definitions,
\beq  \sigm/\sigv\equiv [(M/L\_K)\_{0}/(M\_M/L\_K)]^{1/4} \eeqno{hura}
\par
\begin{figure}[h]
	\centering
\includegraphics[width = 9.5cm] {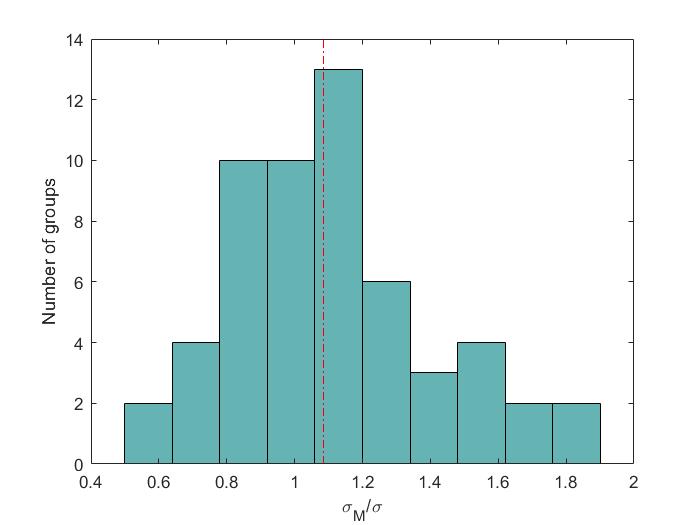}
\caption{The distribution of $\sigm/\sigv$ (on a linear scale). The vertical dash-dotted (red) line marks the median value of 1.085.}		\label{sigmadist}
\end{figure}

Figure \ref{sigmadist} shows the distribution of $\sigm/\sigv$, which has the same content as Fig. (\ref{MONDMoL}), but allows a more direct comparison with the measured line-of-sight dispersions in consideration of the question marks that are known to beset these values (sampling, anisotropies, departure from virialization, inappropriate averaging of measured velocities -- see Sec. \ref{systematics}).
\par
The median of the $\sigm/\sigv$ distribution, which is indicated in various figures below, is clearly $med(\sigm/\sigv)=[med(M\_M/L\_K)]^{-1/4}=1.085$. It is important that it is near 1, but its exact value is not so significant and follows from my choice to calculate $\sigm$ from $L\_K$ using $(M/L\_K)\_0=1\suK$. From eq. (\ref{hura}), the resulting value of $med(\sigm/\sigv)$ scales as $(M/L\_K)\_0^{1/4}$.

\begin{figure}[h]
	\centering
\includegraphics[width = 9.5cm] {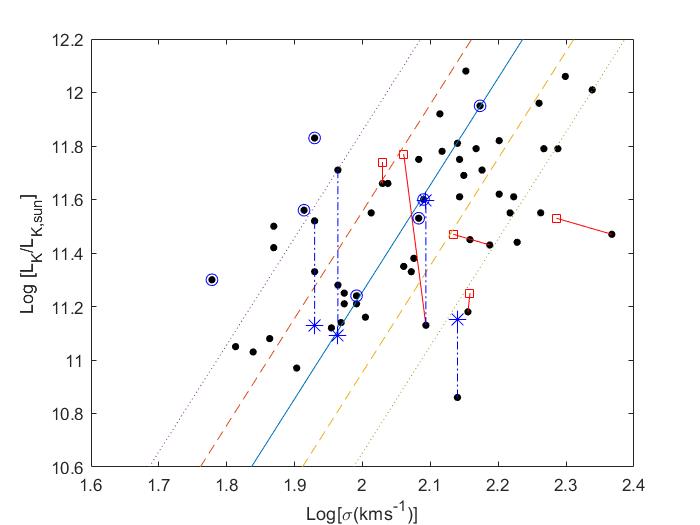}
\caption{$L\_K/L\_{K,\odot}$ against $\sigv$  for values from \Rmk~ (filled circles). The positions  of the five groups from Paper I that pass our cuts are shown as open (red) squares (with $L\_K/L\_{K,\odot}$ and $\sigv$ from Refs. \cite{KNK15} and  \cite{KKN14}). They also appear in our sample, and their positions in the two studies are connected with solid (red) segments. Shown as (blue) asterisks are the distance-corrected positions of the four groups discussed as examples in Sec. \ref{systematics}; they are connected to the uncorrected positions with (blue) dash-dotted segments. The seven additional groups that I checked and that do not seem to require distance correction (see Sec. \ref{systematics})
are surrounded by (blue) circles.
The parallel lines across the plot show the predictions of eq. (\ref{nusa}), $L\_K=(M/L\_K)^{-1}(81/4G\az)\sigv^4$, for the sample's median value, $M/L\_K=0.7\suK$ (solid), for half and for twice this value (dashed, upper and lower, respectively), and for a quarter and four times this value (dotted).}		\label{Lsigma}
\end{figure}

Individual group are represented in the $L\_K-\sigv$ plane in Fig. \ref{Lsigma} (with $L\_K$ and $\sigv$ from \Rmk). Also shown are the MOND predictions of eq. (\ref{nusa}) for several values of the mass-to-light ratios.
\par
Five groups in the sample of Paper I pass our cuts here. They are also in the present sample.
They are ``NGC 3607'', ``NGC 3379'', and ``NGC 3627'', with parameters in Paper I taken from Ref. \cite{KNK15}, and ``NGC 5746'' and ``NGC 5363'' with Paper-I parameters from Ref. \cite{KKN14}. Their positions in the $L\_K-\sigv$ plane are also shown in Fig. \ref{Lsigma} with their parameters from Paper I; they are connected to their positions for the parameters from \Rmk. This comparison gives some notion of the uncertainties in the parameters, which will be discussed in more detail in Sec. \ref{systematics}.
\par
\begin{figure}[h]
	\centering
\includegraphics[width = 9.5cm] {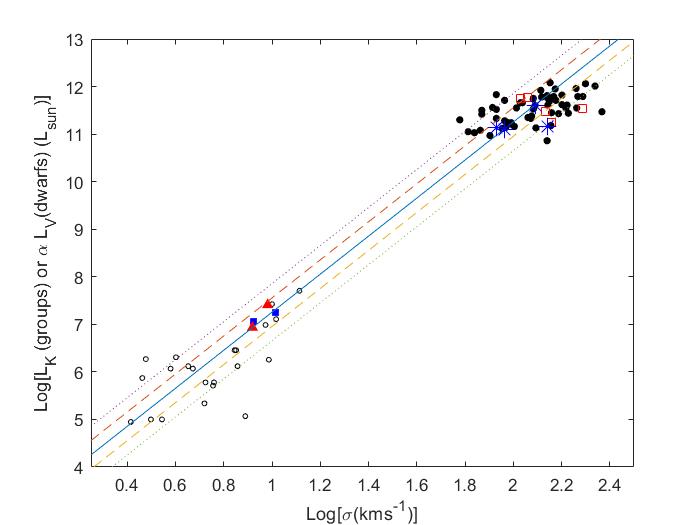}
\caption{$L\_K/L\_{K,\odot}$ for the groups, and $\alpha L\_V/L\_{V,\odot}$ for dwarfs ($\alpha=2/0.7$ corrects for the band difference -- see the reason below), plotted vs. $\sigv$. Groups: filled circles for measured values from \Rmk, open (red) squares for the five groups from Paper I that pass our cuts, and (blue) asterisks are the corrected positions for the four examples discussed in Sec. \ref{systematics} (as in Fig. \ref{Lsigma}). Dwarfs: open circles are for values compiled in Refs. \cite{mm13a,mm13b} with references given therein; filled (blue) squares are for data from Ref. \cite{martin14}; filled (red) triangles are for data from Ref. \cite{kirby14}.  The lines show the predictions of eq. (\ref{nusa}): $L=(M/L)^{-1}(81/4G\az)\sigv^4$.  The solid line is the prediction for $M/L\_K=0.7\suK$ for the groups, and for the more appropriate V-band $M/L\_V=2\suV$ for the dwarfs (hence the upshift of the dwarf $L\_V$ values by the factor $\alpha$). The dashed lines are for 0.5 (upper) and 2 (lower) times these $M/L$ values, and the dotted lines are for 0.25 (upper) and 4 (lower) times these values. Error bars for the dwarfs are given in the literature, but not shown here.}		\label{Lsigmadwarfs}
\end{figure}

The dynamical range of group parameters in our sample is relatively small -- only an order of magnitude in luminosity. To better appreciate the acuteness and significance of the result in Fig. \ref{Lsigma}, I show in Fig. \ref{Lsigmadwarfs} the same plot, but now including the equivalent data for relevant dwarf-spheroidal satellites of Andromeda.\footnote{From Fig. \ref{Lsigmadwarfs} one may get the impression that the slope within the groups is shallower than the value of 4 dictated by MOND. But this is largely an artifact due to our low- and high-luminosity cutoffs.} ``Relevant'' means that they are not clearly dominated by an external-field effect (although some is surely still present), and do not show clear signs of rotation --  either of which would render the use of eq. (\ref{nusa}) invalid. I also took only cases with a large number of observed velocities (when there are two measurements I took the mean of the two), and required that gas does not contribute much to the baryonic mass (otherwise, comparison with stellar $M/L$ ratios is not meaningful). These include 25 dwarfs from the compilations of Refs. \cite{mm13a,mm13b}; two dwarfs with good data from Ref. \cite{martin14}: Cassiopeia III, and Lacerta I;\footnote{These two were already discussed in the context of MOND in Refs. \cite{mm13b,pm14,martin14}. The third in Ref. \cite{martin14}, Per I, has very large velocity errors.} and two dwarfs from Ref. \cite{kirby14}: Cetus and VV 124.

\par
Dwarfs too suffer from systematics that cause artificial scatter, some shared by groups (such as unaccounted-for anisotropies, external-field effects, etc.), some specific to the dwarfs (such as contribution to $\sigv$ from binaries). But I will not dwell on these here, as I am not testing MOND in dwarfs -- this has been done properly in the above references, taking into account rotation, gas content, and possible presence of an external-field effect. Here I only use them as a touchstone for the application of eq. (\ref{nusa}).
\par
\begin{figure}[h]
	\centering
\includegraphics[width = 9.5cm] {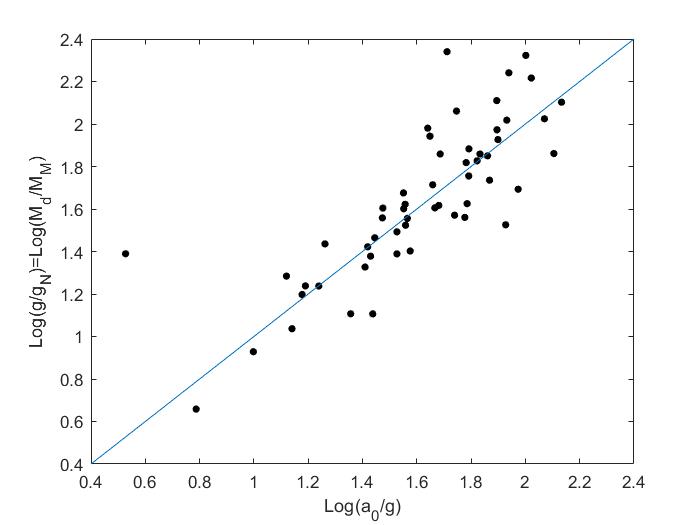}
\caption{The MDAR for the groups in the present sample. The Newtonian-to-MOND mass ratio, which can be viewed as the ratio, $g/\gN$, of the measured acceleration to the Newtonian, baryonic one -- plotted vs. $\az/g$. The equality line -- which the MDAR lies very near to for such low accelerations -- is also shown. }		\label{massratiovsg}
\end{figure}

In light of the discussion in Sec. \ref{aform}, I also show  the results in different forms: Figure \ref{massratiovsg} shows the ratio $M_d/M\_M$ of the Newtonian, dynamical mass to the MOND mass as a function of $\az/g$. This mass ratio can be viewed as ``the mass discrepancy'' in the system, and is also the ratio of the observed acceleration, $g$, to the Newtonian, gravitational acceleration, $\gN$, produced by baryons. Figure \ref{massratiovsg} thus plots the (very low acceleration end of the) MDAR.
This presentation has the advantage that, unlike the $M-\sigv$ plot, it shows the range of accelerations that are probed in our study. We see that the groups satisfy $M_d/M\_M\approx \az/g$, with some scatter, down to values of $g/\az \approx 0.01$, consistent with the expectation from eq. (\ref{mmg}).
\par
\begin{figure}[h]
	\centering
\includegraphics[width = 9.5cm] {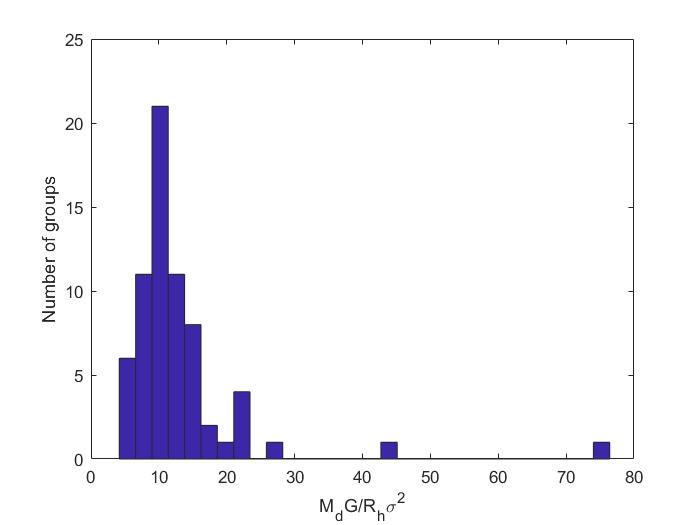}
\caption{The distribution of $M_d G/R_h\sigv^2$ for $M_d$ and $\sigv$ values from \Rmk. }		\label{betlam}
\end{figure}

Note, however, that while our $g$ is defined as in eq. (\ref{gsr}), with $\b=2$ and $\bar R=R_h$,
the values of $M_d$ given in \Rmk~ are not calculated from $\sigv$ and $R_h$, with some constant $\l$, as in our eq. (\ref{mdmd}); they are calculated in some involved way from the individual velocities and projected positions of the members with measured velocities. So, eq. (\ref{mmg}) with these $M_d$ values is not equivalent to the $M\_M-\sigv$ relation.
To see to what extent eq. (\ref{mdmd}) does approximate the $M_d$ values with $\bar R=R_h$, I plot in Fig. \ref{betlam} the distribution of $M_d G /R_h\sigv^2$. We see that the distribution is quite narrow and peaked at $\approx 10$ (the median is 10.6).
Equation (\ref{mmg}) is thus a good approximation with $\bar R=R_h$ and $\b\l^2\approx 10$, which makes the coefficient in eq. (\ref{mmg}) very near 1.
Equation (\ref{mmg}) with $M_d/M\_M\approx \az/g$ is thus a reasonable approximation to the $M\_M-\sigv$ relation, which underlies the behavior shown in Fig. \ref{massratiovsg}.

\par
\begin{figure}[h]
	\centering
\includegraphics[width = 9.5cm] {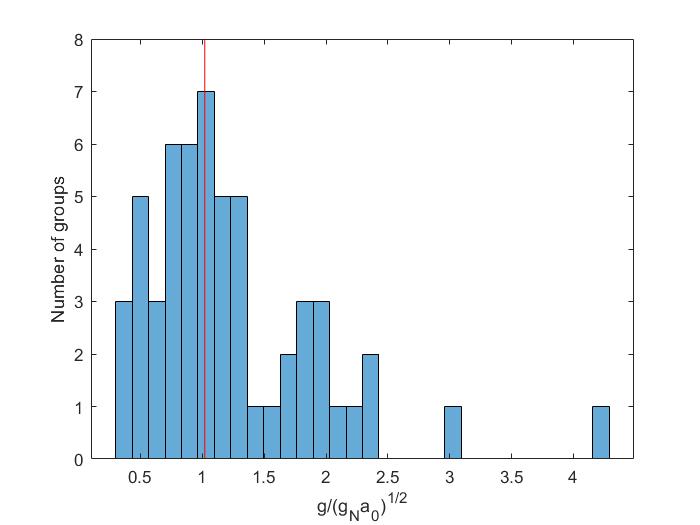}
\caption{The distribution of $g/(\az gN)^{1/2}$. The median value of 1 --  shown as the vertical (red) line -- is built in by the choice of normalization of $\gN$ with factors of order 1.}		\label{histmdar}
\end{figure}

As a more direct comparison with the MDAR in eq. (\ref{mdar}), I show in Fig. \ref{histmdar} a presentation that is {\it  equivalent} to the $M/L\_K$ distribution, i.e. that of $g/(\az \gN)^{1/2}$, where $\gN$ is defined in eq. (\ref{gngn}) with the choice $\k=9/4$ [which makes $\e=1$, in eq. (\ref{mdar}), since $\b=2$], and the baryonic mass determined from $L\_K$, taking the median value of $M/L\_K$ to be $0.7\suK$ that we found for the $M\_M/L\_K$ distribution. A median of 1 is thus enforced with a reasonable values for $\kappa$. This figure shows that the distribution of the ratio is narrow.

\begin{figure}[h]
	\centering
\includegraphics[width = 9.5cm] {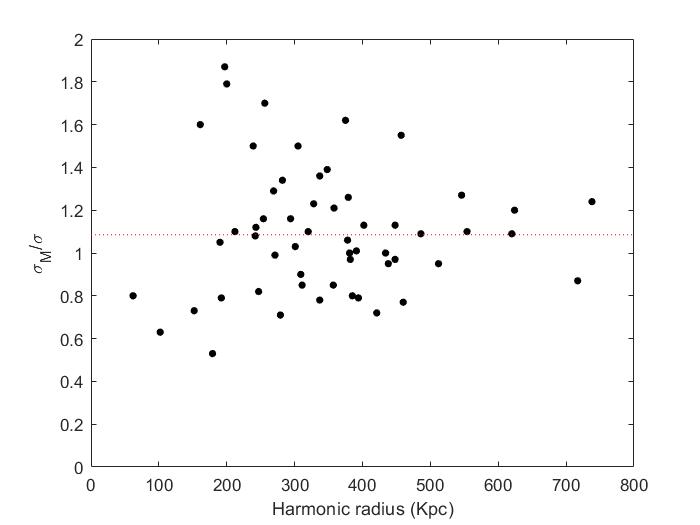}
\caption{$\sigm/\sigv$ plotted against the harmonic radius. The median value of $\sigm/\sigv$ is shown.}		\label{Rhsigma}
\end{figure}

Figure \ref{Rhsigma} shows $\sigm/\sigv$ vs. $R_h$. We see that there is no significant trend with $R_h$. If anything, the agreement with the MOND prediction becomes even tighter at the larger radii, up to $R\_h\approx 700\kpc$.
This is significant in light of the remaining MOND discrepancy in cluster cores, which are not larger in size than the high-$R_h$ groups here. The cluster conundrum is sometimes interpreted as some breakdown of MOND at large scales. The lesson from the groups maybe that it is not the size that matters, but some of the other attributes that differentiate between medium-richness groups and clusters. For example, they have very different velocity dispersions (or depth of the potentials \cite{zf12}). Or, as I discussed in Ref. \cite{milgrom08}, the culprit might be the prevalent hot, x-ray-emitting gas in clusters and rich groups -- absent in large quantities in medium groups -- that makes the difference. The idea is that with such gas comes also a yet-undetected baryonic component, dubbed cluster, baryonic, dark matter. This would account for the remaining MOND discrepancy, and would also account for the observed mass distribution in the ``Bullet Cluster''.

\begin{figure}[h]
	\centering
\includegraphics[width = 9.5cm] {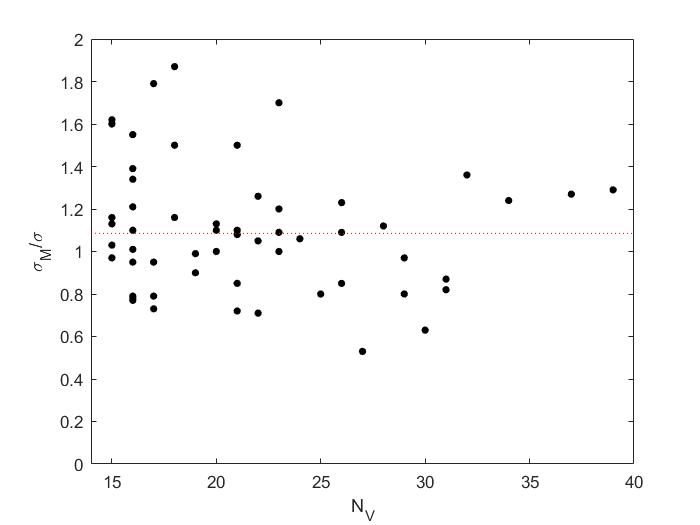}
\caption{$\sigm/\sigv$ (with its median indicated) plotted against the number of members with measured velocities.}		\label{NVsigma}
\end{figure}

Figure \ref{NVsigma} shows $\sigm/\sigv$ vs. $N\_V$, where there is no apparent correlation, in particular no apparent decrease in the scatter with increasing $N\_V$.

\section{Discussion \label{discussion}}
I found that intermediate-richness galaxy groups satisfy the deep-MOND relation $M\_MG\az=(81/4)\sigv^4$, with $M\_M/L\_K$ values that are characteristic of stars. The global dynamics of these very-low-acceleration systems are thus accounted for by MOND with no need for dark matter. This contrasts with the very large quantities of dark matter needed in the framework of Newtonian dynamics -- typically forty times more dark matter than is observed in stars.
\par
This result is accentuated by our reminder that the same MOND relation is obeyed by dwarf spheroidal galaxies.
The groups and dwarfs are two very disparate types of galactic objects, differing by about seven-and-a-half orders of magnitude in baryonic mass, by about two orders in internal velocities, and by three orders in size, and they all lie squarely on the same $M-\sigv$ relation predicted by MOND.
\par
The scatter we see in Fig. \ref{Lsigma} is arguably due partly to genuine variation in the baryonic $M/L\_K$ ratios -- due to scatter in stellar $M/L$ ratios, but also due to the presence of a varying amount of gas (cold and hot), in the member galaxies themselves, and in the intragroup medium, which increases the baryonic values above the stellar values (the mass in the MOND relation is the total baryonic mass). But much scatter surely comes from known (and unknown) systematics, which I discuss below, in Sec. \ref{systematics}.
\par
This scatter is, in any event, much smaller than the factor of about $40$ by which MOND corrects the Newtonian masses.

\subsection{Possible systematics\label{systematics}}
Here I discuss some of the possible systematics that enter the analysis, and the ways they can affect the comparison of the MOND predictions with the observationally deduced quantities.
\par
One important source of error discussed by \Rmk~ is the way they assign distances to the groups.
The distances used by \Rmk~ for deducing their group sizes, luminosities, and dynamical masses, and in turn by me to determine $M\_M/L\_K$, $\sigm$, etc. are not determined directly. They are all based on the observed mean recession velocities with respect to the local group, and assuming a strict Hubble flow, with $H_0=73\kms\mpc^{-1}$. As discussed in \Rmk, this can clearly lead, in some cases, to systematic errors in all the above mentioned system parameters.
To quote \Rmk:
`` The disadvantage of our algorithm, where
the distance of a group is determined by the mean radial velocity
of its galaxies, is most pronounced in the regions with large
peculiar motions. Some groups we identified in the Virgo cluster
core are, most probably, false groups, rather than physical
subsystems in the Virgo cluster. ''
\par
This issue is particularly acute for (nearby) groups with low Hubble velocities -- which \Rmk~ largely excluded from their list. It is also acute when the group resides in a region characterized by high peculiar velocities, such as in a cluster environment, which can modify the recession velocity.
\par
I already mentioned in Sec. \ref{sample} the group ``NGC 4216'', which has a recession speed of $55\kms$, but is ``located in the Virgo cluster core'', and fell prey to this error.
\par
As mentioned in Sec. \ref{results}, five of the groups in the present sample appear in the samples used in Paper I, and pass out cuts. They where assigned different parameter values in the more recent catalogs of Refs. \cite{KNK15,KKN14}, used in Paper I.
These latter assigned distances based on more direct measures.
These differences are evident in Fig. \ref{Lsigma}, where they are compared. We see that the assigned distances (hence $L\_K$) have been updated but also the values of $\sigv$ changed. For example, ``NGC 3379'' has the highest $M\_M/L\_K=12.63\suK$ in our sample (and $\sigm/\sigv=0.53$), which is based on  \Rmk~ giving for it $\sigv=233\kms$ (with $N\_V=27$). The more recent Ref. \cite{KNK15} gives it
$\sigv=193\kms$ (with $N\_V=36$), which, with also an increase in $L\_K$, gives this group $M\_M/L\_K=5.2\suK$ ($\sigm/\sigv=0.66$).
\par
Another of these five groups is the ``NGC 3607'' (Leo II) group. It has a recession speed of $V\_{LG}=960\kms$, which gives a redshift distance of $13\mpc$. Reference \cite{KNK15}, used in Paper I, assigned to it a distance of $25\mpc$, which gave in Paper I, $M\_M/L\_K=0.38\suK$ ($\sigm/\sigv=1.28$). (This shows as the largest shift in Fig. \ref{Lsigma}.)
Moreover, the distance given in Ref.  \cite{brodie14} for this group is $\approx 22.2\mpc$, which would correct the $M\_M/L\_K=2.2\suK$ ($\sigm/\sigv=0.82$) in Table \ref{table1} to $M\_M/L\_K=0.83\suK$ ($\sigm/\sigv=1.05$).
\par
Since the adopted distances may pose such a problem, I checked eleven of the groups to see to what extent distances determined from nonredshift methods (Cepheids, Tully-Fisher, surface-brightness fluctuations, etc.) differ from those adopted in \Rmk. I checked seven outliers and four groups whose $M\_M/L\_K$ are near the median value of $0.7\suK$. I found that for the latter four, nonredshift distances are consistent with the ones adopted, so their near-median $M\_M/L\_K$ is unaltered. Of the seven outliers, three do not seem to require a distance correction (``NGC 3801'', ``NGC 0488'', and ``NGC 891'').
\par
Four of the outliers do require corrections if the nonredshift distances are correct.
One is the above mentioned ``NGC 3607''.
Another, less severe case (not appearing in Paper I), is ``NGC 3031'' (M81), with our second highest $M\_M/L\_K$ value. Based on the recession speed of $V\_{LG}=193\kms$, the redshift-based distance is $2.64\mpc$. But its Cepheid distance is $3.7\mpc$, reducing $M\_M/L\_K$ by a factor of 0.51, from 6.3 in Table \ref{table1} to 3.2 (and corrects to $\sigm/\sigv=0.75$).
\par
Yet another is ``NGC 4254'' (M99), which has $M\_M/L\_K=0.18\suK$. Its recession speed in \Rmk~ is $2296 \kms$, which gives a Hubble-flow distance of $31.45\mpc$.
However, its directly measured distance as given in NASA Extragalactic Database is $D=15.4 \pm 1.7 \mpc$.
With this distance, we should correct to   $M\_M/L\_K=0.75$ (and $\sigm/\sigv= 1.07$).
\par
And yet another is ``NGC 4527'', whose Hubble-flow distance is $D=22\mpc$, but it is in the Virgo cluster in projection; so could have a large peculiar component. Indeed, the Cepheid distance to NGC 4527 (the galaxy) is $14\pm 1.6\mpc$ \cite{saha01}. This requires correcting the MOND $M/L$ value in Table \ref{table1} by a factor 2.5, from the smallish $M\_M/L\_K=0.2\suK$ ($\sigm/\sigv=1.5$) to $M\_M/L\_K=0.5\suK$ ($\sigm/\sigv=1.2$).
\par
All eleven examples are also shown in Fig. \ref{Lsigma}.
\par
While, generally, the Hubble-flow distances are rather reliable, we see that at least some of the scatter must be due to this sometimes-inaccurate distance assignment.
\par
Another source of uncertainty, as mentioned already, is the possible presence in member galaxies or the intragroup space, of additional baryons, such as cold, warm, or hot gas. This may explain some of the higher-than stellar MOND $M\_M/L\_K$ values.
\par
Departure from the velocity isotropy that is assumed in all analyses, when we replace the three-dimensional velocity dispersion by $\sqrt{3}\sigv$, also leads to artificial scatter in the observed $M\_M-\sigv$ relation.
Such anisotropies are surely present, but it is difficult to quantify them without measuring proper motions as well as radial velocities.
\par
In the analyses of whole samples, such as here, such unaccounted-for anisotropies are expected to lead to increased scatter, not to systematic shifts. But one has to be careful not to put too much weight on apparent departures of this or that individual case, be it a group or another ``pressure -supported'' system.
\par
\begin{figure}[h]
	\centering
\includegraphics[width = 9.5cm] {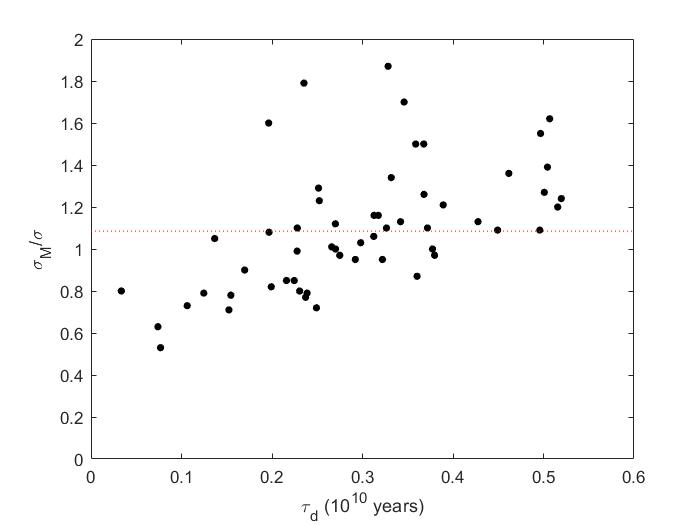}
\caption{$\sigm/\sigv$ plotted against the dynamical time.}		\label{sigmavstau}
\end{figure}

Another possible source of systematics is departure from virialization.
Artificially elevated values of $\sigm/\sigv$ can also be caused, e.g., by the system not yet having reached virialization. I show in Table \ref{table1} a measure of the dynamical time in the group, $\tau_d\equiv R_h/\sigv$. Its values are fractions of the Hubble time, but not exceedingly smaller. It is not clear what is the limit on $\tau_d$ necessary for virialization (and this may anyway vary from system to system).
We expect large dynamical times, of the order of the Hubble time to indicate that the system is still collapsing towards virial equilibrium, and hence has a lower-than-virial value of $\sigv$ resulting in $\sigm/\sigv>1$. We see in Fig. \ref{sigmavstau}, some correlation of high $\sigm/\sigv$ values with high $\tau_d$ values. But it is not clear that this is not an artifact of our definitions, which imply that $\sigm/\sigv\propto \tau_d(L\_K^{1/4}/R_h)$ should hold exactly. So, if, for example, $L\_K^{1/4}/R_h$ is not correlated with $\tau_d$, any scatter in $\sigv$ around the correct value will create such a correlation artificially.

\subsubsection{External-field effect?}
In MOND, an additional source of deviation from relation (\ref{nusa}) is the possible presence of an external-field effect (EFE) due to external accelerations to which the groups are subject (see Refs. \cite{milgrom83,haghi09,mm13a,mm13b} for a few of many treatments of this effect for pressure-supported systems).
\par
The description of the effect is relatively simple only when the (MOND) external acceleration, $g\_{ex}\gg g\_{in}\approx g$, $g\_{in}$ is the (MOND) internal acceleration.
In the opposite case the effect is negligible, and it is rather complicated to account for when $g\_{ex}\approx g\_{in}$, as might be the case for some of our groups.
Equation (\ref{nusa}) -- which assumes that the system is isolated -- underestimates the baryonic mass when an EFE is present, i.e., when $g\_{ex}\not\ll g\_{in}$. It then leads to artificially low $M\_M/L\_K$ values, or high $\sigm/\sigv$ values.\footnote{MOND predicts an acceleration discrepancy $\approx\az/\bar g$, with $\bar g$ the larger of the internal and external accelerations. Assuming isolation in effect uses $\bar g=g\_{in}$ instead of the larger $g\_{ex}$.}
\par
For $g\_{ex}\gg g\_{in}$ the effect works as follows:
Suppose the $\sigv$ value given in \Rmk~ is $\sigv={\sigm}_e$, where ${\sigm}_e$ is the correct dispersion in all regards which would be exactly obtained from a MOND prediction that includes the EFE, using $L\_K$ with a correct $M/L\_K=1\suK$, while ${\sigm}_i$ is what I calculate as $\sigm$ assuming isolation. Then, using eq. (\ref{mdar}) it can be shown that
\beq \frac{{\sigm}_i}{{\sigm}_e}=\e^{1/2}\left(\frac{g\_{ex}}{g}\right)^{1/4}\approx \left(\frac{g\_{ex}}{g}\right)^{1/4}\approx\frac{g\_{ex}^{1/2}}{(\gN\az)^{1/4}}, \eeqno{efeefe}
 where $\e\approx 1$ is defined in eq. (\ref{mdar}), and $g$ is related to $\sigv$ by eq. (\ref{gsr}).
Table \ref{table1} shows the values of $g\equiv 2\sigv^2/R_h$. For many groups $g$ is low enough -- a few percents of $\az$, or even less -- that an EFE may be present due to surrounding structures.  This is because the amplitude of the varying ambient acceleration field may be of that order. It can be estimated, e.g., as the acceleration to typical peculiar velocities of $\approx 300\kms$ during the Hubble time. For example, Ref. \cite{boz18} estimates for the local groups $g\_{ex}\approx 2.2\times 10^{-2}\az$.
\par
If all else was exact, with no measurement errors, and no departure from any of our assumptions, except that of isolation, then departures of $\sigm/\sigv$ from unity would only be due to the EFE, and be given, for $g\_{ex}\gg g$, by eq. (\ref{efeefe}).
Then, if $g\_{ex}$ is not correlated with $g$, we would expect $\sigm/\sigv$ to be unity for $g\ll g\_{ex}$ and broadly behave as $g^{-1/4}$ at very small $g$ values, with scatter that comes from that in $g\_{ex}$ from system to system. We can then look for such a correlation.
\par
However, as in the case of the dynamical time, correlations of $\sigm/\sigv$ with $g$ can also result artificially.
From our definitions in Sec. \ref{aform} we have exactly
\beq \frac{\sigm}{\sigv}=\e^{1/2}\frac{(\az\gN)^{1/4}}{g^{1/2}}.   \eeqno{masut}
In other words, in terms of the quantities from \Rmk~ that we use, and our assumptions that go into the calculation of $\sigm$ and $g$ (e.g., that $M/L\_K=1$) we have, by the definitions, $\sigm/\sigv\propto L\_K^{1/4}R_h^{-1/2}g^{-1/2}$.
So, for example, if there was no EFE, and all measurements were correct except that the $\sigv$ values depart from their correct values,  with no correlation with $\gN$, then eq. (\ref{masut}) tells us to expect $\sigm/\sigv\propto g^{-1/2}$, with scatter reflecting that in $\gN$. This would produce an artificial correlation between $\sigm/\sigv$ and $g$. (Of course, if all quantities are exact, MOND predicts that $g$ is correlated with $\gN$, so that $\sigm/\sigv=1$ for all groups.)
\par
\begin{figure}[h]
	\centering
\includegraphics[width = 9.5cm] {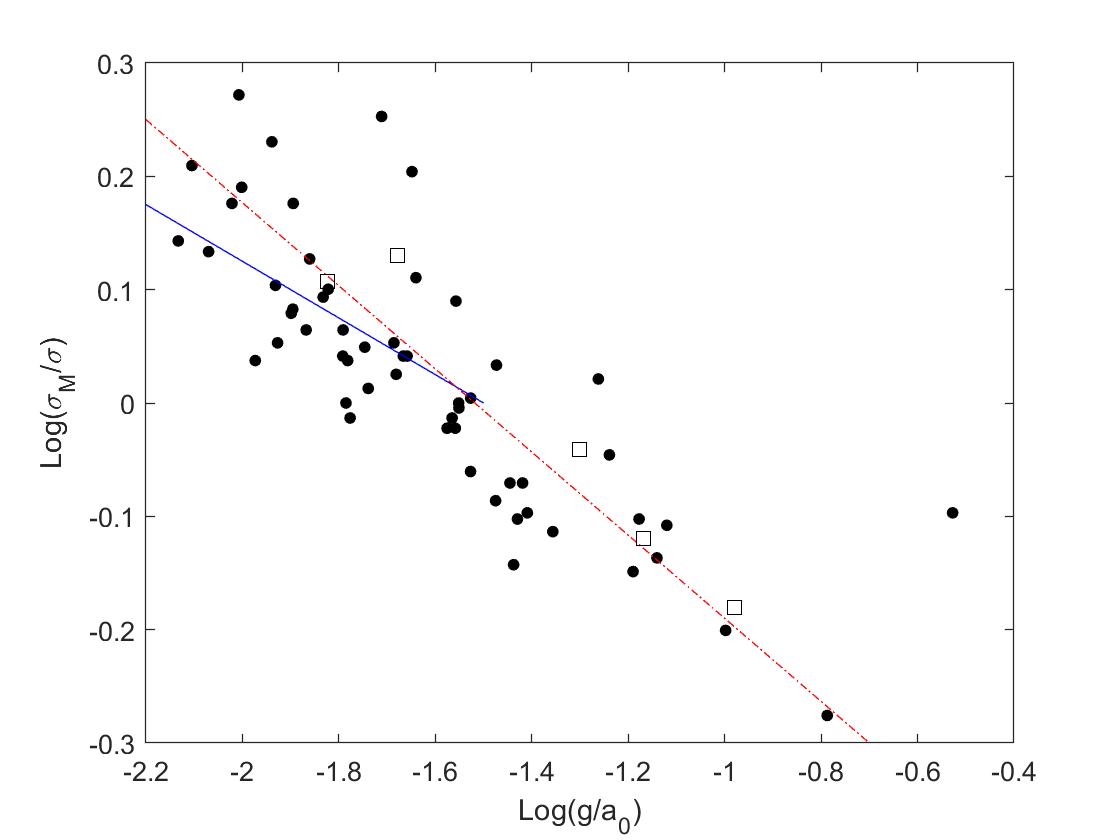}
\caption{$\sigm/\sigv$  plotted against $g/\az$ (filled circles). The position of the five groups from Paper I are shown as squares. The solid (blue) line has a slope of $-1/4$ and follows eq. (\ref{efeefe}) with $g\_{ex}=3\times 10^{-2}\az$. The dash-dotter (red) line has a slope of $-1/2$, as in eq. (\ref{masut}).}		\label{sigmavsg}
\end{figure}

Figure \ref{sigmavsg} shows $\sigm/\sigv$ vs. $g$ for the groups. Also shown are a line of slope $-1/4$ following eq. (\ref{efeefe}), with $g\_{ex}=3\times 10^{-2}\az$, for $g\le g\_{ex}$, and a line with slope $-1/2$ crossing the first line at $g=g\_{ex}$.
\par
 We see clearly that indeed there is a correlation with the unexpectedly high $\sigm/\sigv$ values occurring for the lowest values of $g$, and they decrease with increasing $g$.
However, this correlation follows more closely eq. (\ref{masut}); so much of it may well be an artifact.
It is difficult to judge whether some of the correlation is due to the presence of an EFE.
\par
\begin{figure}[h]
	\centering
\includegraphics[width = 9.5cm] {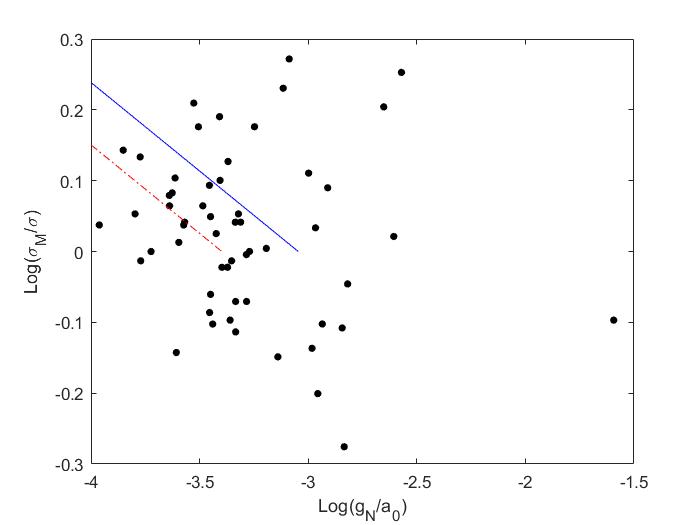}
\caption{$\sigm/\sigv$  plotted against $\gN/\az$ (filled circles).  Lines with a slope of $-1/4$ that follow the {\it asymptotic} ($g\ll g\_{ex}$), last near-equality in eq. (\ref{efeefe}), are also shown with $g\_{ex}=3\times 10^{-2}\az$ (blue, solid) and $g\_{ex}=2\times 10^{-2}\az$ (red, dash dotted).}		\label{sigrvgn}
\end{figure}

Because $\sigv$ enters directly the definition of $g$, which may introduce artificial correlation, it may be more informative to look for a correlation between $\sigm/\sigv$  and $\gN$, as in the last equality in eq. (\ref{efeefe}), since $\gN$ is derived from only $L\_K$ and $R_h$, and does not use $\sigv$.
This is shown in Fig. \ref{sigrvgn}, with the {\it asymptotic} ($g\ll g\_{ex}$) lines of slope $-1/4$ describing the last near equality in eq. (\ref{efeefe}), with $g\_{ex}=(2,~3)\times 10^{-2}\az$. No clear correlation is seen.
\par
It is possible to hide the presence of an EFE in this plot; but unless the scatter in $g\_{ex}$ is very large it seems that most of the scatter in $\sigm/\sigv$ is not caused by an EFE, but by other causes, such as discussed above. In particular, we see that the largest values of $\sigm/\sigv$ do not occur for the lowest $\gN$ groups.

\par
There is a clear potential in the group dynamics to detect the EFE, but in the present analysis, I have not been able to establish its presence, possibly because it is masked by other, more dominant sources of scatter. It is worthwhile investigating this issue further because the EFE is peculiar to MOND, and should not appear in the dark-matter paradigm and could be a discriminating phenomenon. A more thorough analysis may require going beyond statistical arguments and considering individual groups in the context of their environments.

I thank Indranil Banik, Pavel Kroupa, and Stacy McGaugh for comments on the manuscript.
\begin{table*}
\footnotesize
\caption{The 67 groups cataloged in \Rmk, with number of members with measured radial velocities $N\_V\ge 15$ (ordered by decreasing $N\_V$). Groups marked with * are not included in the analysis, for reasons explained in the text. Observed parameters with Newtonian, dynamical masses and mass-to-light ratios as derived in \Rmk~(columns 1-7), and MOND-related quantities that are derived here (columns 8-13).
(1) group name; (2) number of members with measured line-of-sight velocity; (3) line-of-sight velocity dispersion in $\kms$; (4) harmonic radius in \kpc; (5) $\log\_{10}(L\_K/L\_{K,\odot})$; (6) dynamical mass $\log\_{10}(M_d/\msun)$; (7) dynamical $M_d/L\_K$ in solar units; (8) MOND velocity dispersion calculated from eq. (\ref{nusa}), assuming baryonic
$M_b/L\_K=1\suK$, in $\kms$; (9) the ratio $\sigm/\sigv$; (10) MOND mass from $\sigv$ using eq. (\ref{nusa}), in units of $10^{12}\msun$; (11) baryonic, MOND $M\_M/L\_K$ in solar units; (12) a measure of the dynamical time, $\tau_d\equiv R_h/\sigv$, in units of $10^{10}$ years; (13) A measure of the acceleration in the group, $g\equiv 2\sigv^2/R_h$, in units of $10^{-2}\az$
.\label{table1}}
\begin{tabular}{lcccccccccccc}

\hline
\multicolumn{1}{c}{Group}   &   $N\_V$& $\sigv$&   $R_h$&    $\lg L\_K$ &  $\lg M_d$  &  $M_d/L\_K$ & $\sigm$ & $\sigm/\sigv$
& $M\_{M,12}$ &  $M\_M/L\_K$ & $\tau_{d,10}$& \multicolumn{1}{c}{${\small 10^2}g/\az$}
\\
\hline
\multicolumn{1}{c}{(1)}& \multicolumn{1}{c}{(2)}& (3) &
\multicolumn{1}{c}{(4)} & \multicolumn{1}{c}{(5)}&
\multicolumn{1}{c}{(6)} &\multicolumn{1}{c}{(7)} & (8)
&\multicolumn{1}{c}{(9)} &(10)& (11) & (12) &(13)
\\
\hline
\hline
    NGC4472* &  355 &  291  &  696  &  12.44  &  14.14  &  50  &  216  &  0.74  &  9.07  &  3.29  &  0.24  &  6.55       \\
    NGC3311* &  139 &  426  &  520  &  12.51  &  14.29  &  60  &  225  &  0.53  &  41.66  &  12.87  &  0.12  &  18.78       \\
    NGC4696* &  116 &  303  &  690  &  12.5  &  14.13  &  43  &  224  &  0.74  &  10.66  &  3.37  &  0.23  &  7.16       \\
    NGC1316* &  111 &  244  &  454  &  12.3  &  13.94  &  44  &  200  &  0.82  &  4.48  &  2.25  &  0.19  &  7.06       \\
    NGC4261* &  87  &  276  &  358  &  11.99  &  13.7  &  51  &  168  &  0.62  &  7.34  &  7.51  &  0.13  &  11.45       \\
    NGC5846* &  74  &  228  &  395  &  11.8  &  13.64  &  69  &  150  &  0.66  &  3.42  &  5.42  &  0.17  &  7.08       \\
    NGC3992* &  72  &  120  &  452  &  11.68  &  13.33  &  45  &  140  &  1.16  &  0.26  &  0.55  &  0.38  &  1.71       \\
    NGC5371* &  55  &  195  &  455  &  12.07  &  13.69  &  42  &  175  &  0.9  &  1.83  &  1.56  &  0.23  &  4.5       \\
    NGC3223* &  53  &  404  &  368  &  12.13  &  14.31  &  151  &  181  &  0.45  &  33.7  &  24.98  &  0.09  &  23.86       \\
    NGC5044* &  52  &  245  &  480  &  11.96  &  13.72  &  58  &  164  &  0.67  &  4.56  &  5  &  0.2  &  6.73       \\
    NGC5746  &  39  &  107  &  269  &  11.66  &  13.2  &  35  &  138  &  1.29  &  0.17  &  0.36  &  0.25  &  2.29       \\
    NGC4697  &  37  &  109  &  546  &  11.66  &  13.27  &  41  &  138  &  1.27  &  0.18  &  0.39  &  0.5  &  1.17       \\
    NGC3100  &  34  &  142  &  738  &  12.08  &  13.57  &  31  &  176  &  1.24  &  0.51  &  0.43  &  0.52  &  1.47       \\
    NGC4636  &  32  &  73   &  337  &  11.08  &  12.58  &  32  &  99  &  1.36  &  0.04  &  0.3  &  0.46  &  0.85       \\
    NGC3607  &  31  &  124  &  247  &  11.13  &  13.08  &  89  &  102  &  0.82  &  0.3  &  2.22  &  0.2  &  3.35       \\
    NGC4501  &  31  &  199  &  717  &  12.06  &  13.79  &  54  &  174  &  0.87  &  1.98  &  1.73  &  0.36  &  2.97       \\
    NGC3031  &  30  &  138  &  102  &  10.86  &  12.59  &  54  &  87  &  0.63  &  0.46  &  6.33  &  0.07  &  10.04       \\
    NGC1553  &  29  &  185  &  62  &  11.79  &  13.56  &  59  &  149  &  0.8  &  1.48  &  2.4  &  0.03  &  29.7       \\
    NGC4105  &  29  &  139  &  382  &  11.61  &  13.23  &  42  &  134  &  0.97  &  0.47  &  1.16  &  0.27  &  2.72       \\
    NGC4631  &  28  &  90   &  243  &  11.12  &  12.98  &  72  &  101  &  1.12  &  0.08  &  0.63  &  0.27  &  1.79       \\
    NGC3379  &  27  &  233  &  179  &  11.47  &  13.23  &  58  &  124  &  0.53  &  3.73  &  12.63  &  0.08  &  16.32       \\
    NGC3923  &  26  &  159  &  357  &  11.62  &  13.33  &  51  &  135  &  0.85  &  0.81  &  1.94  &  0.22  &  3.81       \\
    ESO507--025&26  &  130  &  328  &  11.92  &  13.18  &  18  &  160  &  1.23  &  0.36  &  0.43  &  0.25  &  2.77       \\
    NGC5078  &  26  &  138  &  620  &  11.81  &  13.48  &  47  &  151  &  1.09  &  0.46  &  0.71  &  0.45  &  1.65       \\
    NGC1407  &  25  &  167  &  385  &  11.61  &  13.32  &  51  &  134  &  0.8  &  0.98  &  2.42  &  0.23  &  3.9       \\
    NGC1395  &  24  &  121  &  378  &  11.53  &  13.05  &  33  &  128  &  1.06  &  0.27  &  0.8  &  0.31  &  2.08       \\
    NGC4039  &  23  &  74   &  256  &  11.5  &  12.82  &  21  &  126  &  1.7  &  0.04  &  0.12  &  0.35  &  1.15       \\
    NGC4303  &  23  &  115  &  434  &  11.35  &  12.97  &  42  &  116  &  1  &  0.22  &  0.99  &  0.38  &  1.64       \\
    NGC4535  &  23  &  121  &  624  &  11.75  &  13.36  &  41  &  145  &  1.2  &  0.27  &  0.48  &  0.52  &  1.26       \\
    NGC4753  &  23  &  98   &  486  &  11.21  &  12.76  &  35  &  107  &  1.09  &  0.12  &  0.72  &  0.5  &  1.06       \\
    NGC0988  &  22  &  103  &  379  &  11.55  &  12.98  &  27  &  130  &  1.26  &  0.14  &  0.4  &  0.37  &  1.51       \\
    NGC1332  &  22  &  183  &  279  &  11.55  &  13.39  &  69  &  130  &  0.71  &  1.42  &  4  &  0.15  &  6.46       \\
    NGC7176  &  22  &  139  &  190  &  11.75  &  13.11  &  23  &  145  &  1.05  &  0.47  &  0.84  &  0.14  &  5.47       \\
    NGC4373  &  21  &  149  &  554  &  11.95  &  13.4  &  28  &  163  &  1.1  &  0.62  &  0.7  &  0.37  &  2.16       \\
    NGC5322  &  21  &  169  &  421  &  11.44  &  13.12  &  48  &  122  &  0.72  &  1.03  &  3.75  &  0.25  &  3.65       \\
    NGC3877  &  21  &  65   &  239  &  11.05  &  12.57  &  33  &  97  &  1.5  &  0.02  &  0.2  &  0.37  &  0.95       \\
    NGC3894  &  21  &  123  &  242  &  11.6  &  13.02  &  26  &  133  &  1.08  &  0.29  &  0.73  &  0.2  &  3.36       \\
    NGC2911  &  21  &  144  &  311  &  11.45  &  13.2  &  56  &  122  &  0.85  &  0.54  &  1.93  &  0.22  &  3.59       \\
    NGC4111  &  20  &  93   &  212  &  11.14  &  12.69  &  35  &  102  &  1.1  &  0.09  &  0.69  &  0.23  &  2.19       \\
    NGC5011  &  20  &  131  &  448  &  11.78  &  13.43  &  45  &  148  &  1.13  &  0.37  &  0.62  &  0.34  &  2.06       \\
    NGC5557  &  20  &  141  &  381  &  11.69  &  13.3  &  41  &  141  &  1  &  0.5  &  1.02  &  0.27  &  2.81       \\
    NGC3610  &  19  &  119  &  271  &  11.38  &  13.08  &  50  &  118  &  0.99  &  0.25  &  1.06  &  0.23  &  2.81       \\
    NGC6868  &  19  &  182  &  309  &  11.96  &  13.38  &  26  &  164  &  0.9  &  1.39  &  1.52  &  0.17  &  5.77       \\
    NGC0891  &  18  &  60   &  197  &  11.3  &  12.64  &  22  &  112  &  1.87  &  0.02  &  0.08  &  0.33  &  0.98       \\
    NGC4527  &  18  &  85   &  305  &  11.52  &  12.93  &  26  &  127  &  1.5  &  0.07  &  0.2  &  0.36  &  1.27       \\
    NGC5473  &  18  &  94   &  294  &  11.25  &  12.75  &  32  &  109  &  1.16  &  0.1  &  0.56  &  0.31  &  1.62       \\
    NGC0488  &  17  &  85   &  200  &  11.83  &  13.16  &  21  &  152  &  1.79  &  0.07  &  0.1  &  0.24  &  1.94       \\
    ESO320--031&17  &  150  &  438  &  11.71  &  13.33  &  42  &  142  &  0.95  &  0.64  &  1.25  &  0.29  &  2.76       \\
    NGC5363  &  17  &  143  &  152  &  11.18  &  12.76  &  38  &  105  &  0.73  &  0.53  &  3.49  &  0.11  &  7.24       \\
    NGC4321  &  17  &  165  &  394  &  11.55  &  13.35  &  63  &  130  &  0.79  &  0.94  &  2.64  &  0.24  &  3.72       \\
    NGC0524  &  16  &  147  &  391  &  11.79  &  13.16  &  23  &  149  &  1.01  &  0.59  &  0.96  &  0.27  &  2.97       \\
    NGC3627  &  16  &  154  &  192  &  11.43  &  13.05  &  42  &  121  &  0.79  &  0.71  &  2.64  &  0.12  &  6.65       \\
    NGC3945  &  16  &  92   &  358  &  11.28  &  12.93  &  45  &  111  &  1.21  &  0.09  &  0.48  &  0.39  &  1.27       \\
    NGC4125  &  16  &  85   &  282  &  11.33  &  12.67  &  22  &  114  &  1.34  &  0.07  &  0.31  &  0.33  &  1.38       \\

 \hline
\end{tabular}
\end{table*}

\addtocounter{table}{-1}
\begin{table*}
\footnotesize
\caption{Continue}
\begin{tabular}{lcccccccccccc}
\hline
\multicolumn{1}{c}{Group}   &   $N\_V$& $\sigv$&   $R_h$&    $\lg L\_K$ &  $\lg M_d$  &  $M_d/L\_K$ & $\sigm$ & $\sigm/\sigv$
& $M\_{M,12}$ &  $M\_M/L\_K$ & $\tau_{d,10}$& \multicolumn{1}{c}{${\small 10^2}g/\az$}
\\
\hline
\multicolumn{1}{c}{(1)}& \multicolumn{1}{c}{(2)}& (3) &
\multicolumn{1}{c}{(4)} & \multicolumn{1}{c}{(5)}&
\multicolumn{1}{c}{(6)} &\multicolumn{1}{c}{(7)} & (8)
&\multicolumn{1}{c}{(9)} &(10)& (11) & (12) &(13)
\\
\hline
\hline

    NGC4151  &  16  &  69   &  348  &  11.03  &  12.56  &  34  &  96   &  1.39  &  0.03  &  0.27  &  0.5   &  0.74       \\
    NGC4216* &  16  &  52   &  23   &  8.6    &  11.24  &  437 &  24   &  0.46  &  0.01  &  23.23 &  0.04  &  6.33       \\
    NGC4254  &  16  &  92   &  457  &  11.71  &  13.28  &  37  &  142  &  1.55  &  0.09  &  0.18  &  0.5   &  1          \\
    NGC4666  &  16  &  98   &  320  &  11.24  &  12.95  &  51  &  108  &  1.1   &  0.12  &  0.67  &  0.33  &  1.61       \\
    NGC4936  &  16  &  194  &  460  &  11.79  &  13.36  &  37  &  149  &  0.77  &  1.79  &  2.91  &  0.24  &  4.4        \\
    NGC5090  &  16  &  218  &  337  &  12.01  &  13.74  &  54  &  169  &  0.78  &  2.86  &  2.79  &  0.15  &  7.59       \\
    NGC5982  &  16  &  159  &  512  &  11.82  &  13.31  &  31  &  151  &  0.95  &  0.81  &  1.22  &  0.32  &  2.66       \\
    NGC3801  &  15  &  82   &  161  &  11.56  &  12.7   &  14  &  130  &  1.6   &  0.06  &  0.16  &  0.2   &  2.25       \\
    NGC4224  &  15  &  118  &  448  &  11.33  &  12.95  &  42  &  114  &  0.97  &  0.25  &  1.15  &  0.38  &  1.67       \\
    NGC4258  &  15  &  80   &  254  &  10.97  &  12.45  &  30  &  93   &  1.16  &  0.05  &  0.56  &  0.32  &  1.36       \\
    NGC4993  &  15  &  74   &  375  &  11.42  &  12.44  &  10  &  120  &  1.62  &  0.04  &  0.14  &  0.51  &  0.79       \\
    NGC5128  &  15  &  94   &  402  &  11.21  &  12.52  &  20  &  107  &  1.13  &  0.1   &  0.61  &  0.43  &  1.18       \\
    NGC5198  &  15  &  101  &  301  &  11.16  &  12.69  &  34  &  104  &  1.03  &  0.13  &  0.91  &  0.3   &  1.82       \\

\hline
\end{tabular}
\end{table*}



\end{document}